\documentclass[12pt]{article}

\def\be{\begin{equation}}
\def\ee{\end{equation}}
\def\bea{\begin{eqnarray}}
\def\eea{\end{eqnarray}}

\title{\bf Detectability of cosmic topology in almost f\/lat 
universes}

\author{
G.I. Gomero$^1$\thanks{gomero@cbpf.br}, \ 
M.J. Rebou\c{c}as$^1$\thanks{reboucas@cbpf.br}, \  
R. Tavakol$^{1,2}$\thanks{r.tavakol@qmw.ac.uk} \\ 
\\ 
$^{1}$  Centro Brasileiro de Pesquisas F\'\i sicas, \\
Rua Dr. Xavier Sigaud 150 \\
22290-180 Rio de Janeiro -- RJ, Brazil\\
\\
$^{2}$ Astronomy Unit, School of Mathematical Sciences, \\
Queen Mary, University of London, \\
Mile End Road, London E1 4NS, UK
}

\begin{document}

\date{\today}

\maketitle

\begin{abstract} \noindent
Recent observations suggest that the ratio 
of the total density to the critical density of the universe, 
$\Omega_0$, is likely to be very close to one, with a signif\/icant 
proportion of this energy being in the form of a dark component with 
negative pressure. 
Motivated by this result, we study the question of observational 
detection of possible non-trivial topologies in universes
with $\Omega_0 \sim 1$, which include a cosmological constant.
Using a number of indicators we f\/ind that as $\Omega_0 \to 1$, 
increasing families of possible manifolds (topologies) become either 
undetectable or can be excluded observationally.
Furthermore, given a non-zero lower bound on $\left|\Omega_0 -1\right|$,
we can rule out families of topologies (manifolds) as possible candidates 
for the shape of our universe. We demonstrate these f\/indings 
concretely by considering families of manifolds and 
employing bounds on cosmological parameters from recent 
observations. We f\/ind that given the present bounds on 
cosmological parameters, there are families of both hyperbolic 
and spherical manifolds that remain undetectable and
families that can be excluded as the shape of our universe.
These results are of importance in future search strategies for 
the detection of the shape of our universe, given that there are
an inf\/inite number of theoretically possible topologies and 
that the future observations are expected to put a non-zero 
lower bound on $\left|\Omega_0 - 1 \right|$ which is more 
accurate and closer to zero.
\end{abstract}

\section{Introduction}

A number of recent observations suggest that the ratio of the 
baryonic (plus dark) matter density to the critical density, 
$\Omega_{m0}$, is signif\/icantly less than unity, being of the 
order of $\Omega_{m0} \sim 0.2-$0.3 (see e.g.%
~\cite{ostriker-steinhardt95} and references therein).
On the other hand, recent measurements of the position of the 
f\/irst acoustic peak in the angular power spectrum of CMBR 
anisotropies, by BOOMERANG-98 and MAXIMA-I experiments, seem 
to provide strong evidence that the corresponding ratio 
for the total density to the critical density, $\Omega_0$, is close 
to one~\cite{boomerang1}~--~\cite{maxima1}. 
Added to this is the  evidence from the spectral and photometric 
observations of Type Ia Supernovae~\cite{Schmidt,Riess1}
which seem to suggest that the universe is undergoing an accelerated 
expansion at the present epoch~\cite{perl,perl1}. 
This rather diverse set of observations has led to an  
evolving consensus among cosmologists that the total density
of the universe, $\Omega_0$, is likely to be very close 
to unity, 
a signif\/icant part of which resides 
in the form of a dark component which 
is smooth on cosmological scales and which possesses negative
pressure. A candidate that satisf\/ies these criteria
is the vacuum energy corresponding to a cosmological 
constant. 

Somewhat parallel to these developments, and to some extent
unrelated to them, a great deal of work has also recently gone 
into studying the possibility that the universe may possess compact 
spatial sections with a non-trivial topology (see, for example,
~\cite{LeUzLu}~--~\cite{ZelNov83}),
including the construction of dif\/ferent topological indicators
(see e.g.~\cite{SokShv74}~--~\cite{GRT01}). 
A fair number of these studies have concentrated 
on cases where the densities corresponding to matter and vacuum 
energy are substantially smaller than the critical density.
This was reasonable because until very 
recently observations used to point to a low density universe. 
In addition an important aim of most of these previous works has 
often been to produce examples where the topology of the universe 
has strong observational signals and can therefore be detected 
and even determined.

The main aims of this paper, which are complementary to these 
earlier works, are twofold. Firstly, to study the question of 
detectability of a possible non-trivial compact topology in locally
homogeneous and isotropic universes whose total density 
parameter is taken to be close to one ($\Omega_0 \sim 1$). 
Secondly, to show how given a non-zero lower bound on  
$\left |\Omega_0 - 1 \right|$, we can exclude certain families 
of compact manifolds as viable candidates for the shape of 
the universe. We shall do this by employing two indicators, 
namely the ratios of the so called {\em injectivity radius\/} 
($\,r_{inj}\,$), and {\em maximal inradius\/} ($\,r_{-}^{max}\,$) 
to the depth $\,\chi_{obs}\,$ of given catalogues. 
We study almost f\/lat models with both $\Omega_0>1$ and 
$\Omega_0<1$, and our results apply to any method of detection 
of topology based on observations of multiple images of either 
cosmic objects or spots of microwave background radiation.
We note that even though the idea that non-trivial
topologies become harder to detect as $\Omega_0 \to 1$ is 
implicitly present in some other works~\cite{CSS98a,LeLuUz},
it has, however, often been passed over as an uninteresting limit
concerning non-trivial topologies, given the observations 
at the time.
Here, in addition to studying this limit in detail
by considering concrete families of manifolds (topologies),
rather than individual examples as has often been done before,
we treat it as the most relevant limit concerning
the geometry of the universe today in view of the recent 
observations~\cite{boomerang1}~--~\cite{maxima1}.

The structure of the paper is as follows. In Section~\ref{CosSet} 
we give an account of the cosmological models employed. 
Section~\ref{Top} contains a brief account of our topological 
setting. In Section~\ref{DetTop} we discuss the 
question of detectability of cosmic topology using a number
of indicators. Section~\ref{AlmostFlat} contains a detailed 
discussion of the question of detection of cosmic topology
in universes with $\left|\Omega_0 -1 \right| \ll 1$, with 
the help of concrete examples, and f\/inally 
Section~\ref{Concl} contains summary of our main results
and conclusions. 

\section{Cosmological Setting} \label{CosSet}

Let us begin by assuming that the universe is modelled by a 
$4$-manifold $\mathcal{M}$ which allows a $(1+3)$ splitting, 
$\mathcal{M} = R \times M$, with a locally isotropic and 
homogeneous Friedmann-Lema\^{\i}tre-Robertson-Walker (FLRW) 
metric
\be
\label{FLRW1}
ds^2 = -c^2dt^2 + R^2 (t) \left [ d \chi^2 + f^2(\chi) (d\theta^2 + 
\sin^2 \theta  d\phi^2) \right ]\; ,
\ee
where $t$ is a cosmic time, the function $f(\chi)$ is given by 
$f(\chi)= \chi\,,\;$ $\sin\chi\,,\;$ or $\sinh\chi\,,\;$ 
depending on the sign of the constant spatial curvature 
($k = 0, \pm 1$), and $R(t)$ is the scale factor.
Furthermore, we shall assume throughout this article that 
the $3$-space $M$ is a multiply connected compact
quotient manifold of the form $M = \widetilde{M} /\Gamma$, 
where $\Gamma$ is a discrete group of isometries of 
$\widetilde{M}$ acting freely on the covering space 
$\widetilde{M}$ of $M$, where $\widetilde{M}$  
can take one of the forms $E^3, S^3$ or $H^3$ [corresponding, 
respectively, to f\/lat ($k=0$), elliptic ($k>0$) and 
hyperbolic ($k<0$) spaces]. The group $\Gamma$ is called 
the covering group of $M$, and is isomorphic to its 
fundamental group $\pi_1(M)$.

For non-f\/lat models ($k \neq 0$), the scale factor 
$R(t)$ is identif\/ied with the curvature radius of the 
spatial section of the universe at time $t$, and thus $\chi$ can 
be interpreted as the distance of any point with coordinates 
$(\chi, \theta, \phi)$ to the origin of coordinates 
(in $\widetilde{M}$), in units of curvature radius,
which is a natural unit of length 
and suitable for measuring areas and volumes. 
Throughout this paper we shall use this natural unit.

In the light of current observations, we assume the current 
matter content of the universe to be well approximated by 
dust (of density $\rho_m$) plus a cosmological constant 
$\Lambda$. The Friedmann equation is then given 
by
\be
H^2 =\frac{8 \pi G \rho_m }{3} -\frac{k c^2}{R^2} 
            +\frac{\Lambda}{3}\;,
\ee
where $H=\dot{R}/R$ is the Hubble parameter and $G$ is 
the Newton's constant, which upon
introducing $\Omega_m = \frac{8 \pi G \rho_m}{3 H^2}$ and 
$\Omega_{\Lambda} \equiv \frac{8 \pi G \rho_{\Lambda}}{3 H^2} = 
\frac{\Lambda}{3 H^2}$  and letting
$\Omega = \Omega_m + \Omega_{\Lambda}$, it can 
be rewritten as 
\be   \label{R-eq}
R^2 = \frac{k c^2}{ H^2 \left( \Omega - 1 \right) } \;.
\ee

For  universe models having compact spatial sections with 
non-trivial topology, which we shall be concerned with 
in this article, it is clear that any attempt at
the discovery of such a topology through observations must start 
with the comparison of the curvature radius and the 
horizon radius (distance) $d_{hor}$ at the present time. 
To calculate the latter, we recall the
redshift-distance relation in the above FLRW settings can
be written in the form 
\begin{equation}
\label{redshift-dist}
d(z) = \frac{c}{H_0} \int_0^z \left[ (1+x)^3 \Omega_{m0} + 
\Omega_{\Lambda 0} - (1+x)^2 (\Omega_0 -1) \right]^{-1/2} dx \; ,
\end{equation}
where the index $0$ denotes evaluation at present time.
The horizon radius $d_{hor}$ is then def\/ined as
\begin{displaymath}
d_{hor} = \lim_{z \to \infty} d(z) \; ,
\end{displaymath}
which, by using (\ref{R-eq}) evaluated at the present time, 
can be expressed in units of the curvature radius in the 
form
\begin{equation} \label{horLambda}
\chi_{hor} \equiv \frac{d_{hor}}{R_0} =\sqrt{|1-\Omega_0|} 
   \int_0^{\infty} \left[ (1+x)^3 \Omega_{m0} + 
   \Omega_{\Lambda 0} - (1+x)^2 (\Omega_0 -1) \right]^{-1/2} dx \; .
\end{equation}

We note that in practice there are basically three 
types of catalogues which can be used in order to search 
for repeated patterns in the universe and hence the nature of 
the cosmic topology: namely, clusters of galaxies, containing
clusters with redshifts of up to $z \approx 0.3$;
active galactic nuclei (mainly QSO's and quasars),
with a redshift cut-of\/f of $z_{max} \approx 4$; and 
the CMBR from the 
decoupling epoch with a redshift of $z \approx 3000$.
It is expected that future catalogues will increase the 
f\/irst two cut-of\/fs to $z_{max} \approx 0.7$~\cite{CNOC2} and 
$z_{max} \approx 6$ (by 2005~\cite{SDSS}) respectively.
Thus, instead of $\chi_{hor}\,$ (for which  $z \to \infty\,$) 
it is observationally more appropriate to consider 
the largest distance $\chi_{obs} = \chi(z_{max})$ 
explored by a given astronomical survey. 

\section{Topological prerequisites} \label{Top}

Let us start by recalling that, contrary to the hyperbolic and 
elliptic cases, there is no natural 
unit of length for the Euclidean geometry, 
since in this case the curvature radius is inf\/inite. Also,
it is unlikely that astrophysical observations can 
f\/ix the density parameter $\Omega_0$ to be exactly equal to one. 
Therefore, in this section we shall conf\/ine ourselves to nearly 
f\/lat hyperbolic and spherical spaces and brief\/ly consider 
some relevant facts about manifolds with constant non-zero 
curvature, which will be used in the following sections.

Regarding the hyperbolic case, there is at present no complete 
classif\/ication of these manifolds. A number of important results 
are, however, known about them, including the two important theorems 
of Mostow~\cite{Mostow} (see also Prasad~\cite{Prasad}) and 
Thurston~\cite{Thurston97}.
According to the former, geometrical quantities of orientable 
hyperbolic manifolds, such as their (f\/inite) volumes and the 
lengths of their closed geodesics, are topological invariants. 
Moreover, it turns out that there are only a f\/inite number 
of compact hyperbolic manifolds with the same volume.
According to the latter, there is a countable 
inf\/inity of sequences of compact orientable hyperbolic manifolds,
with the manifolds of each sequence being ordered in terms of their 
volumes~\cite{Thurston97}. 
Moreover there exists a hyperbolic 3-manifold with 
a minimum volume~\cite{Thurston97}, whose volume is shown to be 
greater than 0.28151~\cite{Przeworski}.

Hyperbolic manifolds can be easily constructed and studied with 
the publicly available software package SnapPea~\cite{SnapPea} 
(see e.g.~\cite{AdamsSnap}). In this way, Hodgson and Weeks have 
compiled a census of 11031 closed hyperbolic manifolds, ordered 
by increasing volumes (see~\cite{SnapPea,HodgsonWeeks}). 
As examples, we show in Table~\ref{Tb:HW-Census} the f\/irst 
10 manifolds from this census with the lowest volumes,
together with their volumes, as well as the lengths $l_M$ of their 
smallest closed geodesics and their inradii $\,r_{-}\,$, which
are discussed below. The f\/irst two entries 
(rows) of Table~\ref{Tb:HW-Census} are known as Weeks' 
and Thurston's manifolds respectively, with the former 
conjectured to be the smallest (minimum volume) closed hyperbolic 
manifold. 
%
\begin{table}[!htb]
\begin{center}
\begin{tabular}{*{4}{|c}|} \hline
$M$ & $Vol(M)$ & $l_M$ & $r_-$  \\ \hline \hline  
 m003(-3,1) & 0.942707 & 0.584604 & 0.519162 \\ \hline
m003(-2,3) & 0.981369 & 0.578082 & 0.535437 \\ \hline
 m007(3,1) & 1.014942 & 0.831443 & 0.527644 \\ \hline
 m003(-4,3) & 1.263709 & 0.575079 & 0.550153 \\ \hline
 m004(6,1) & 1.284485 & 0.480312 & 0.533500 \\ \hline
m004(1,2) & 1.398509 & 0.366131  & 0.548345 \\ \hline
 m009(4,1) & 1.414061 & 0.794135 & 0.558355  \\ \hline
m003(-3,4) & 1.414061 & 0.364895 & 0.562005 \\ \hline
 m003(-4,1) & 1.423612 & 0.352372 & 0.535631 \\ \hline
 m004(3,2) & 1.440699 & 0.361522 & 0.556475 \\ \hline
\end{tabular}
\caption[Hodgson-Weeks census.] {\label{Tb:HW-Census} 
\small First ten manifolds in the Hodgson-Weeks
census of closed hyperbolic manifolds, together with their 
corresponding volumes, length $l_M$ of their smallest closed 
geodesics and their inradii $r_{-}$.
}
\end{center}
\end{table}

Regarding the spherical manifolds, we recall that the isometry 
group of the $3$-sphere is $O(4)$. However, since any isometry
of $S^3$ that has no f\/ixed points is orientation-preserving,
to construct multiply connected spherical manifolds it is
suf\/f\/icient to consider the subgroups of $SO(4)$, because for
these subgroups the orientation is necessarily preserved.
Thus any spherical 3-manifold of positive constant curvature 
is of the form $S^3/\Gamma$, where $\Gamma$ is a f\/inite 
subgroup of $SO(4)$ acting freely on the $3$-sphere. 
The classif\/ication of spherical $3$-dimensional manifolds is well 
known and can be found, for instance in~\cite{Wolf,Scott,Thurston82} 
(see also~\cite{Ellis71} for a description in the context of cosmic 
topology). In the remainder of this section we shall
focus our attention on the important subset of such
spaces, referred to as {\em lens spaces\/},
in order to examine the question of detectability of
cosmic topology of nearly f\/lat spherical universes.

Brief\/ly, there are an inf\/inite number of 3-dimensional 
lens spaces that are globally homogeneous and also an inf\/inite 
number that are {\em only\/} locally homogeneous. 
In both cases the lens spaces are quotient spaces of the form 
$S^3/Z_p$ where the covering groups are the cyclic groups $Z_p$ 
($p \geq 2$). 
The cyclic groups $Z_p$ can act on $S^3$ in dif\/ferent ways 
parameterized by an integer parameter, denoted by $q$, such 
that $p$ and $q$ are relatively prime integers such that 
$1\! \leq q < p/2\,$. These quotients are the lens spaces 
$L(p,q)$. 
The action of $Z_p$ on $S^3$ gives rise to  globally homogeneous 
lens spaces only if $q = 1$; in all other cases the quotient 
$S^3/Z_p$ gives rise to lens spaces that are globally inhomogeneous.

One can give a very simple description of the actions 
of $Z_p$ on $S^3$ that give rise to lens spaces. In fact, the 
3-sphere can be described as the points of the bi-dimensional 
complex space $C^2$ with modulus 1, thus
\begin{displaymath}
S^3 = \left\{ (z_1,z_2) \in C^2 \; : \; |z_1|^2 + |z_2|^2 = 1 
\right\} \, ,
\end{displaymath}
and any action of $Z_p$ on $S^3$ giving rise to a lens space 
is generated by an isometry of the form
\begin{eqnarray}
\label{lensspace}
\alpha_{(p,q)} : S^3 & \rightarrow & S^3 \nonumber \\
           (z_1,z_2) & \mapsto     & \left( e^{2 \pi i/p} z_1, 
e^{2 \pi i q/p} z_2 \right) \,,
\end{eqnarray}
where $p$ and $q$ are given as above.

To close this section it is important to recall that since 
any covering group of spherical space forms is of f\/inite order, 
all of its elements are cyclic. As a consequence, any 3-dimensional 
spherical manifold is f\/initely covered by a lens space. Thus 
lens spaces play a special role in studying the topology of 
spherical manifolds, which
motivates their employment below in our study of the 
problem of detectability of the cosmic topology in nearly 
f\/lat spherical universes.

\section{Detectability of cosmic topology} \label{DetTop}

Unless there are fundamental laws that restrict the topology 
of the universe, its detection and determination is ultimately 
expected to be an observational problem, at least at a classical 
level. As a f\/irst step, according to eq.~(\ref{horLambda}),
one would expect the topology of nearly f\/lat compact
universe to be detectable from cosmological
observations, provided the bound $\chi_{hor} \ge 1\,$  
($d_{hor} \ge R_0$) holds, since the typical lengths of the 
simplest hyperbolic and spherical manifolds are of the order of 
their curvature radii (see, for example, the values of 
$l_M$ and $r_-$ in Table~\ref{Tb:HW-Census}).

This is, however, a rough estimate and a more appropriate bound will 
crucially depend upon the detailed shape of the universe, our position 
in it, and on the  cosmological parameters. This is particularly
true in view of the fact that a crucial feature of generic 3-manifolds 
is their complicated shapes. Moreover, the fact that most 3-manifolds
with non-zero curvature are globally inhomogeneous introduces the 
possibility of {\em observer dependence\/} (or {\em location 
dependence}) in the topological indicators, in the sense made precise 
below, which makes it likely for these bounds to be location dependent.
What is therefore called for are indicators which are more
sensitive (accurate) than $\chi_{hor}$ and which at the same time take 
into account the uncertainty in our location in such compact 
universes.

A natural way to characterize the shape of compact manifolds
is through the size of their closed geodesics.
More precisely, for any $x \in M$, the distance function $\delta_g(x)$ 
for a given isometry $g \in \Gamma$ is def\/ined by
\be
\label{dist-function}
\delta_g(x) = d(x,gx) \; ,
\ee
which gives the length of the closed 
geodesic associated with the isometry $g$ that passes 
through $x$. So one can def\/ine the length of the smallest 
closed geodesic that passes through $x \in M$ as
\begin{displaymath}
\ell(x) = \min_{g \in \widetilde{\Gamma}} \{\delta_g(x)\} \; ,
\end{displaymath}
where $\widetilde{\Gamma}$ denotes the covering group without the 
identity map, i.e. $\widetilde{\Gamma} = \Gamma \setminus \{id\}$.  

In a globally homogeneous manifold, the distance function for 
any covering isometry $g$ is constant, as is the
length of the closed geodesic associated with it.
However, this is not the case in a locally, but non-globally, 
homogeneous manifold, where in general the length of the smallest 
closed geodesic in $M$ is given by%
\footnote{We are assuming that the topology of the spacelike 
section $M$ of our universe is compact, so for any possible 
manifold a closed geodesic of non-zero minimum length always 
exists.}
\be
\label{smallest-CG}
l_{M} = \inf_{x \in M} \{\ell(x)\} \; = \!\!
\inf_{(g,x) \in \widetilde{\Gamma} \times M} \{\delta_g(x)\} \;,
\ee
where $\inf_{x \in M} \{\ell(x)\}$ denotes the absolute (global) 
minimum of $\ell(x)\,$,
and $l_M$ clearly satisf\/ies $l_{M} \leq \ell(x)$.
The {\em injectivity radius\/}, which is the radius of the
smallest sphere inscribable in
$M$ (terminology used in the SnapPea
package~\cite{SnapPea} and in refs.~\cite{LuRou,CSS98a}) 
is then given by $r_{inj} = l_{M}/2$. 
One can similarly def\/ine the injectivity radius at 
any point $x$ as $r_{inj} (x) = \ell(x) /2$.

Furthermore, the {\em maximal inradius\/} $r_{-}^{max}$, can be 
def\/ined as the radius of the largest sphere embeddable in $M$, 
and is given by 
\begin{displaymath}
r_{-}^{max} = \frac{1}{2} \, \sup_{x \in M} \{\ell(x)\} \; ,
\end{displaymath}
where $\sup_{x \in M} \{\ell(x)\}$ indicates the absolute (global) 
maximum of $\ell(x)\,$. 
The maximal inradius $\,r_{-}^{max}$ is half of the largest distance 
any point in $M$ can be from its closest image.
Clearly in the covering space $\widetilde{M}$, the maximal inradius 
$\,r_{-}^{max}$ is also the radius of the largest sphere inscribable 
in any fundamental polyhedron (FP) of the set of all possible 
fundamental polyhedra of $M$. 

An indicator that has been utilized in most studies regarding 
searches for topological multiple images, is the ratio of the 
injectivity radius $r_{inj}$ to $\chi_{hor}$ (see, for example,
~\cite{LuRou} and reference therein and also~\cite{gert} for a 
similar measure). 
At f\/irst sight this seems to be a very accurate indicator, since 
it def\/ines the minimum scale required for multiple images to be 
in principle observable in a given multiply connected universe.  
This can be made more practical from an observational point of view
by taking instead a dif\/ferent indicator, def\/ined as the ratio of 
the injectivity radius to the largest distance explored by some 
astronomical survey $\chi_{obs}\,$~\cite{LuRou}, namely
\be 
\label{r_inj}
T_{inj}=\frac{r_{inj}}{\chi_{obs}} \; .
\ee

It turns out, however, that such individual indicators based
on topological invariants do not encode all the information 
required to f\/ix the topology uniquely. 
Thus, more than one indicator is often necessary in 
practice. This is the case for the above indicator 
$T_{inj}$, an important limitation of which arises from 
the fact that generic (globally inhomogeneous) manifolds 
are likely to have complicated shapes, making it
unlikely for the smallest closed geodesic to pass precisely 
through our location in the universe.  
There is also the fact that in the set of all compact manifolds
there is no lower bound to the length of the smallest 
closed geodesics  (or equivalently on $r_{inj}$), even though 
each given manifold does have a lower bound $l_{M}\,$.

To partially remedy the f\/irst shortcoming above, one can consider,
for an observer ${\cal O} (x)$ situated at $x \in M$, the analogous 
location dependent measure 
\be
T_{{inj}} (x) =\frac{r_{inj} (x) }{\chi_{obs}} \;.
\ee
However, there is still a major problem associated with this 
indicator, since our uncertainty regarding our location in the 
universe makes $r_{inj} (x)$ uncertain.

As another indicator, we take the ratio of the length 
of the {\em maximal inradius\/} to $\chi_{obs}\,$, namely
\be 
\label{r_-max}
T_{max}=\frac{r_{-}^{max}}{\chi_{obs}} \;.
\ee
A crucial feature of this indicator is that it provides a bound 
which holds for all observers, regardless of their location
and in this sense, therefore, circumvents the problem of our 
ignorance regarding our location in the universe. We note also that 
assuming a f\/ixed astronomical survey for a given compact universe, 
$T_{inj} (x)$ is bounded by the values of $r_{inj}$ and 
$r_{-}^{max}\,$, thus
\be
T_{inj} \le T_{inj} (x) \le T_{max}\;. 
\ee

It should be noted that one cannot guarantee that the inradius 
(given in the fourth column of Table~\ref{Tb:HW-Census}) is the 
maximal inradius for the corresponding hyperbolic manifolds. This
is so because the inradius is calculated for a specif\/ic Dirichlet 
domain, which depends on the basepoint used for its construction. 
Thus, the same manifold can have dif\/ferent Dirichlet 
domains and so dif\/ferent inradii. The maximal inradius 
$\,r_{-}^{max}$ of a given manifold $M$, however, is unique%
~\cite{Weeks} and corresponds to the largest sphere inscribable in 
any possible fundamental polyhedra of $M\,$. 
The package SnapPea chooses the basepoint $x$ 
at a local maximum of the inradius, which may or may not be the 
global (absolute) maximum $r_{-}^{max}$, but the relation 
$\,r_{-}^{max} \geq r_-\,(x)\,$ clearly holds for all compact 
hyperbolic (and elliptic) manifolds. We shall show in the next 
section that the indicator $T_{max}$ will enable us to exclude 
certain families of manifolds (topologies) as the shape of our 
universe.

Let us close this section by computing $r_{inj}$ and $r_{-}^{max}$ 
for the lens spaces which will be used in the following sections. 
Recall that the distance between two points 
$z=(z_1,z_2)$ and $w=(w_1,w_2)$ on $S^3$ is the angle between 
these points viewed from the origin of complex bi-dimensional
space $C^2$,
\begin{displaymath}
\cos \left( d(z,w) \right) =\langle z,w \rangle 
                           = Re(z_1w_1^* + z_2w_2^*) \; ,
\end{displaymath}
where $\langle \, , \rangle$ indicates the inner product in 
$C^2$, and $Re$ takes the real part of any complex expression. 
Taking $w = \alpha_{(p,q)} z$, with $\alpha_{(p,q)}$ as 
in~(\ref{lensspace}), one has
\begin{equation}
\label{distlens}
\cos \left( d(z,w) \right) = \cos \frac{2 \pi}{p} 
      + \left( \cos \frac{2 \pi q}{p} 
      - \cos \frac{2 \pi}{p} \right) |z_2|^2 \; .
\end{equation}
If $q = 1$, then $d(z,w) = 2 \pi/p$ is position independent, 
showing that $L(p,1)$ is globally homogeneous, as anticipated in 
the previous section. In the general case of a lens space $L(p,q)$ 
one obtains
\begin{equation}
\label{r_inj-lens}
r_{inj} =\frac{\pi}{p} \qquad \mbox{and}
                       \qquad r_{-}^{max} = \frac{\pi q}{p},
\end{equation}
since the coef\/f\/icient of 
$|z_2|^2$ in (\ref{distlens}) is always non-positive.

\section{Detectability of topology of nearly f\/lat 
universes} \label{AlmostFlat}

In this section we study how the bounds provided by recent 
cosmological observations can constrain the set of detectable 
topologies and how given a non-zero lower bound on
$\left |\Omega_0 - 1 \right |$, we can exclude certain 
families of manifolds as viable candidates for the shape 
of our universe.

To study the constraints on detectability as a function of 
$\Omega_0$, we begin by considering the horizon radius%
~(\ref{horLambda}). 
Figures~1a and 1b show the behaviour of $\chi_{hor}$ as a 
function of $\Omega_{m0}$ and $\Omega_{\Lambda 0}$.
Since in $\Omega_{m0}\,$--$\,\,\Omega_{\Lambda 0}$ 
plane (hereafter referred to as the parameter space) the
f\/lat universes are obviously characterized by the straight line 
$\Omega_{m0}+\Omega_{\Lambda 0}=1$, these f\/igures clearly 
demonstrate that as $|\Omega_0 -1| \to 0$ then $\chi_{hor} \to 0$, 
hence showing that, for a given manifold $M$ with non-zero curvature, 
there are values of $|\Omega_0 -1|$ below which the topology 
of the universe is undetectable ($\chi_{hor} <1)$ for any 
mix of $\Omega_{m0}$ and $\Omega_{\Lambda 0}$. 
A crucial feature of this result is the rapid way $\chi_{hor}$ 
drops to zero  in a narrow neighbourhood of the 
$\Omega_0 = 1$ line. From the observational point of view,
this is signif\/icant since it shows that the  
detection of the topology of the nearly f\/lat 
models ($\Omega_0 \sim 1$) becomes more and more dif\/f\/icult
as $\Omega_0 \to 1$, a limiting value
favoured by recent observations.

To show concretely how $T_{inj}$ and $T_{max}$ can be used to set 
constraints on the topology of the universe, we recall 
two recent sets of bounds on cosmological parameters%
~\cite{nu2K,Bond-et-al-00a}, namely 
\begin{itemize}
\item[({\bf i})]
$\Omega_0 = 1.08 \pm 0.06$, and $\Omega_{\Lambda 0} = 0.66
\pm 0.06$, obtained by combining the data from
CMBR and galaxy clusters; and 
\item[({\bf ii})]
 $\Omega_0 = 1.04 \pm 0.05$, and
$\Omega_{\Lambda 0} = 0.68 \pm 0.05$, obtained
by combining the CMBR, supernovae and large scale structure 
observations.
\end{itemize}
The bound ({\bf i}) exclusively implies 
positive curvatures for the spatial 3-surfaces, while
({\bf ii}) allows negative curvatures as well. We note that 
even though the precise values of these bounds are likely to
be modif\/ied by future observations, the closeness of 
$\Omega_0$ to $1$ is expected to be conf\/irmed.
We have chosen these bounds here as concrete examples of
how recent observations may be employed in order to
constrain the topology of the universe, and clearly similar
procedures can be used for any modif\/ied future bounds on
cosmological parameters.

Let us return to the question of detectability of the topology 
in the neighbourhood  $\Omega_0 \sim 1$ and assume that 
a particular catalogue which covers the entire sky up to a 
redshift cut-of\/f $z_{max}$.
For a given universe 
[with f\/ixed ($\Omega_{m0}\,$,$\,\,\Omega_{\Lambda 0}\,$)] 
this assumption f\/ixes the values of $\chi_{obs}$ used 
in (\ref{r_inj})~--~(\ref{r_-max}).

Suppose that the universe is compact and let $r_{inj}$ 
denote its injectivity radius. Now if 
$\,\chi_{obs} < r_{inj}\,$, then the topology of the universe 
is undetectable by any survey of depth up to the corresponding 
$z_{max}\,$. 
Note that this assertion holds regardless of our particular 
position in the universe, as $r_{inj} \leq r_{inj}(x),$ for 
all $x$. Also, since $\chi_{hor}$ is the farthest 
distance at which events are causally connected to us,
one can obtain limits on undetectability by 
considering $\chi_{hor}$ rather than $\chi_{obs}$.
For example, for a given universe [f\/ixed 
($\Omega_{m0}\,$,$\,\,\Omega_{\Lambda 0}$)] any manifold $M$ 
whose $r_{inj}$ has a value that lies above the bird-like 
surface in Figure~1 (thus ensuring $T_{inj} > 1$) is 
undetectable. 
The crucial point here is that as we approach the line
$\Omega_0 = 1$, the rapid decrease in the allowed values
of $\chi_{obs}$ will result in a rapid elimination of
families of detectable manifolds (topologies). 
This is important given that recent observations seem
to restrict the cosmological parameters to small regions 
near the $\Omega_0=1$ line.

As concrete examples of constraints set by the bound ({\bf i}), 
we consider universes that possess lens space $L(p,q)$ topologies.
Using~(\ref{r_inj-lens}), and recalling that the topology
cannot be detected if $\,T_{inj} >1\,$, one f\/inds that
for this family of universes the topology  
will be undetectable if 
\be
\label{lens-und}
p < \frac{\pi}{\chi_{obs}} \leq p_* \, ,
\ee
where $p_*$ is the smallest integer larger than $\pi/\chi_{obs}\,$.
Using (\ref{horLambda}) and (\ref{lens-und}), we have compiled in 
Table~\ref{Tb:lens-und} the values of $p_*$ 
for dif\/ferent sets of values of ($\Omega_0$, $\Omega_{\Lambda 0}$) 
contained in bounds ({\bf i}) and four catalogues with distinct
values of $z_{max}\,$. 
%
\begin{table}[!htb]
\begin{center}
\begin{tabular}{*{5}{|c}|} \hline
$\; z_{max} \;$ & $\;\; \Omega_0 \;\;$ & $\;\; \Omega_{\Lambda 0} 
\;\;$ & $\chi_{obs}$ & $\;\; p_* \;\;$ \\ \hline \hline
         & 1.02 & 0.60 & 0.40885 &  8 \\ \cline{2-5}
$\infty$ & 1.08 & 0.66 & 0.82728 &  4 \\ \cline{2-5}
         & 1.14 & 0.72 & 1.10777 &  3 \\ \hline \hline
         & 1.02 & 0.60 & 0.40088 &  8 \\ \cline{2-5}
  3000   & 1.08 & 0.66 & 0.81135 &  4 \\ \cline{2-5}
         & 1.14 & 0.72 & 1.08670 &  3 \\ \hline \hline
         & 1.02 & 0.60 & 0.24375 & 13 \\ \cline{2-5}
    6    & 1.08 & 0.66 & 0.49596 &  7 \\ \cline{2-5}
         & 1.14 & 0.72 & 0.66796 &  5 \\ \hline \hline
         & 1.02 & 0.60 & 0.10278 & 31 \\ \cline{2-5}
    1    & 1.08 & 0.66 & 0.20879 & 16 \\ \cline{2-5}
         & 1.14 & 0.72 & 0.28073 & 12 \\ \hline
\end{tabular}
\caption[Undetectable lens-spaces.] {\label{Tb:lens-und}
\small 
For each value of $z_{max}$ the f\/irst and third rows correspond 
to the smallest and largest values of $\chi_{obs}$ in the 
parameter space given by bounds ({\bf i}), while the second entry 
corresponds to the central value. From this table one can see 
that the projective space $L(2,1)$, and the lens space $L(3,1)$ 
are  undetectable in a universe with $\Omega_0 = 1.08$, 
and $\Omega_{\Lambda 0} = 0.66$; while for the same values of
the density parameters the lens spaces $L(4,1)$, 
$L(5,1)$, $L(5,2)$, and $L(6,1)$ are undetectable, using 
catalogues of depth up to 
$z_{max}=6$. }
\end{center}
\end{table}

As  concrete examples, we note that given $\Omega_0 = 1.08$ and 
$\Omega_{\Lambda 0} = 0.66$, then according to Table~\ref{Tb:lens-und}, 
it would be impossible to detect any 
multiple images if the universe turns out to be the projective 
3-space $L(2,1)\,$ (for which $r_{inj}=  1.57080\,$), or the lens 
space $L(3,1)\,$ (for which $r_{inj}=  1.04712\,$), as the entire 
observable universe would lie inside some fundamental polyhedron of $M$. 
Similarly, if the universe turns out to be either of the
lens spaces $L(4,1)$, $L(5,1)$, $L(5,2)$, or $L(6,1)$, it 
would be impossible to detect its topology using catalogues 
of quasars up to $z_{max}=6$. 
It is important to note that from Table~\ref{Tb:lens-und},
one can see that as $\Omega_0$ approaches unity, $p_*$ increases,
which implies that an increasing subset of lens space topologies 
are undetectable, in agreement with Figures~1.
Note also that the greater the depth $z_{max}\,$ of survey, the 
smaller is the number of undetectable topologies, as expected.
But even when $z_{max} \to \infty $
there is still a subset of lens space topologies
that remains undetectable.
The key point here is that proceeding in a similar way,
one can translate bounds on cosmological parameters to
constraints on allowed topologies (here taken to be lens 
topologies).

So far we have considered the question of detectability of
families of manifolds (topologies) in universes with $\Omega_0 \to 1\,$.
Alternatively, we may ask what is the region of the parameter 
space for which a given topology is undetectable. To this end,
we note that for a given topology (f\/ixed $r_{inj}$)
and for a given catalogue cut-of\/f $z_{max}$, one can solve the 
equation 
\be
\label{largeUniv}
\chi_{obs} = r_{inj} \,,
\ee
which is an implicit function of $\Omega_{m0}$ and 
$\Omega_{\Lambda 0}$. Equivalently using (\ref{horLambda}),
equation~(\ref{largeUniv}) can be rewritten as the 
following implicit function of $\varepsilon_0 \equiv 1-\Omega_0$ 
and $\varepsilon_\Lambda \equiv 1-\Omega_{\Lambda 0}\,\,$:
\begin{equation}
\label{epseq}
\chi_{obs} \equiv \sqrt{|\varepsilon_0|} \int_0^{z_{max}} 
 \left[ (\varepsilon_{\Lambda} - \varepsilon_0)(1+x)^3 
+1-\varepsilon_{\Lambda} +\varepsilon_0(1+x)^2 \right]^{-1/2} dx\,
            =\, r_{inj} \;.
\end{equation}

Now, note that any point of the parameter space for 
which $\chi_{obs} < r_{inj}$, will lie below the graph of the 
solution of eq.~(\ref{epseq}) in the plane 
$(\varepsilon_{\Lambda}, |\varepsilon_0|)$. Thus the points 
below the solution curves correspond to universe models for 
which any topology with injectivity radius larger than 
that given by the f\/ixed $r_{inj}$ are undetectable.

As concrete examples of the constraints set by the bound 
({\bf ii}), we consider the subinterval of this bound 
with $\Omega_0 \leq 1\,$, corresponding to nearly flat 
hyperbolic universes. We will take $r_{inj}$ as the
largest value of $\,\chi_{obs}\,$ in this region of parameter 
space, which can be easily calculated to be $\chi_{obs} = 0.20125$ 
for $z_{max} = 6\,$, and $\chi_{obs} = 0.34211$ for $z_{max} = 3000$.
We solved equation~(\ref{epseq}) with these two specif\/ic  
extreme values of $\chi_{obs}$ as the values for $r_{inj}\,$, and
the results are shown in Fig.~2 as plots of 
$\varepsilon_0$ against $\varepsilon_{\Lambda}$. As can be 
seen the allowed hyperbolic region of the parameter space,
given by $\varepsilon_0 \in (0, 0.01]$ and 
$\varepsilon_{\Lambda} \in [0.27,0.37]$, lies
below both curves, showing  that using quasars up to $z_{max}=6\,$, 
nearly f\/lat FLRW hyperbolic universes, with the density parameters 
in this region, will have undetectable topologies, 
if their corresponding injectivity radii satisfy
$r_{inj} \geq  0.20125\,$. 
Similarly, using CMBR, the  topology of nearly f\/lat FLRW 
hyperbolic universes with $r_{inj} \geq 0.34211$ and the 
density parameter in the hyberbolic range of the bound ({\bf ii}) 
are undetectable.

In Table~\ref{Tb:HW-Census1} we have summarized the restrictions 
on detectability imposed by the hyperbolic subinterval of the 
bounds ({\bf ii}) on the f\/irst seven manifolds of Hodgson-Weeks 
census of closed orientable hyperbolic manifolds (there is
no restrictions for the last three in the set of the
f\/irst ten smallest manifolds).
In Table~\ref{Tb:HW-Census1}, $\,U$ denotes that the topology 
is undetectable by any survey of depth up to the redshifts 
$z_{max}=6$ (quasars) or  $z_{max}=3000$ (CMBR) respectively. Thus
using quasars, the topology of the f\/ive known smallest 
hyperbolic manifolds, as well as m009(4,1), are undetectable 
within the hyperbolic region of the parameter space given
by ({\bf ii}),
while only topologies  m007(3,1) and m009(4,1) 
remain undetectable even if CMBR observations are used.
%
\begin{table}[!ht]
\begin{center}
\begin{tabular}{*{4}{|c}|} \hline
$M$ & $r_{inj}$ & {\sc quasars} & {\sc cmbr}  \\ \hline \hline 
 m003(-3,1) & 0.292302 & $U$ & --- \\ \hline
m003(-2,3) & 0.289041 & $U$ & --- \\ \hline
 m007(3,1) & 0.415721 & $U$ & $U$ \\ \hline
 m003(-4,3) & 0.287539 & $U$ & --- \\ \hline
 m004(6,1) & 0.240156 & $U$ & --- \\ \hline
m004(1,2) & 0.183065 & ---  & --- \\ \hline
 m009(4,1) & 0.397067 & $U$ & $U$  \\ \hline
\end{tabular}
\caption{ \label{Tb:HW-Census1} 
\small 
Restrictions on detectability imposed by the hyperbolic interval 
of the recent bounds ({\bf ii}) for the f\/irst seven manifolds 
of Hodgson-Weeks census. The capital $U$ stands for undetectable
using catalogues of quasars (up to $z_{max}=6$) or
CMBR (up to $z_{max}=3000$).
}
\end{center}
\end{table}

Hitherto we have used $T_{inj}$ together with observational bounds 
on cosmological parameters in order to set bounds on detectability.
We shall now employ the indicator $T_{max}$ 
as a way of excluding certain families of
manifolds (topologies) for the universe. 
Recalling that for a hyperbolic or spherical compact manifold $M$, 
the maximal inradius $r_{-}^{max}$ is the radius of the 
largest sphere embeddable in $M$, then  any catalogue of depth 
$z_{max}$, such that $\chi_{obs} > r_{-}^{max}$, may contain 
multiple images of cosmic sources or CMBR spots. 
Thus if we can be conf\/ident that multiple images do not 
exist up to a certain depth $z_{max}$, then one can claim 
that $T_{max} > 1\,$ ($r_{-}^{max} > \chi_{obs}$), and therefore
any topology not satisfying this inequality (i.e. those
for which $r_{-}^{max} < \chi_{obs}\,$) can be excluded by such 
observations.%
\footnote{Obviously it is possible that even exploring a region 
of the universe deeper than a ball of radius $r_{-}^{max}$ with 
catalogues of cosmic sources, we do not observe any multiple images. 
This may be the case when the selection rules used to
build the catalogues do not allow the recording of enough
multiple images to have a detectable topological signal 
(see~\cite{GTRB98} for a more detailed discussion on this point).}

As an illustration for the case of spherical manifolds, 
suppose again that our universe has a lens space topology. 
If using a catalogue of cosmic objects  
with redshift cut-of\/f up to $z_{max}$ yields no multiple 
images, then any lens space satisfying 
$T_{max}\! < 1\,$ ($r_{-}^{max} < \chi_{obs}$) can be discarded 
as a model for the shape of our universe, 
while those with $T_{max}\! > 1\,$ would be either detectable or
undetectable. 
Using~(\ref{r_inj-lens}) one can see that the discarded lens 
spaces $L(p,q)$ are those satisfying
\be
\label{lens-discard}
p \geq p_q^* \, ,
\ee
where $p_q^*$ is the smallest integer larger than $q\pi/\chi_{obs}\,$.
As concrete examples, assume that a catalogue of cluster of 
galaxies with redshift cut-of\/f $z_{max}=1$ was constructed
and no multiple images exist. Using the values of $\chi_{obs}\,$
{}from Table~\ref{Tb:lens-und}, corresponding to the extremes 
and the central values of $\Omega_0$ and $\Omega_{\Lambda 0}$ 
of the bounds ({\bf i}), we can 
obtain values of $p_q^*\,$ corresponding to $q$ ranging from $1$ to $7$,
as shown
in Table~\ref{Tb:lens-discard}.
%
\begin{table}[!hbt]
\begin{center} \begin{tabular}{|c|c||c|l|l|l|l|l|l|l|} 
\hline 
$\Omega_0$ & $\Omega_{\Lambda 0}$ &  $q$  $\;\rightarrow$ & 
         1 & 2 & 3 & 4 & 5 & 6  &  7    \\ \hline \hline
$1.14$ & $0.72$ & $p^*_q \,\rightarrow$ &
12 & 23 & 34  &  45 & 56  & 68  & 79            \\ \hline 
$1.08$ & $0.66$ & $p^*_q \,\rightarrow$ &
16 & 31 & 46  & 61  & 76  & 91  & 106           \\ \hline
$1.02$ & $0.60$ & $p^*_q \,\rightarrow$ &
31 & 62 & 93  & 123 & 154 & 185 & 216           \\ 
\hline 
\end{tabular} \end{center} 
\caption[]{\label{Tb:lens-discard} 
\small Lens spaces $L(p,q)$ with $p \geq p_q^*$ are discarded on 
the basis of observations if there are no multiple images up to 
$z_{max} = 1$ in universes with three dif\/ferent set of
($\Omega_0\,,\,\Omega_{\Lambda 0}$) in the bound ({\bf i}). 
For a given universe [f\/ixed ($\Omega_0\,,\,\Omega_{\Lambda 0}$)] 
and for each value of $q$ the corresponding value of $p_q^*$ is 
given, thus for example, any lens space of the family $L(p,3)$ 
with $p \geq 93$ could not be the shape of a universe with
$\Omega_0=1.02$ and $\Omega_{\Lambda 0}=0.60\,$.}
\end{table}
Equation~(\ref{lens-discard}) together with this table make it
clear that, for example, any lens space belonging to the 
family $L(p,4)$, with $p \geq 61$, cannot be the shape of an
elliptic universe with the central values of the density 
parameters.
Furthermore, it is clear from Table~\ref{Tb:lens-discard} 
that as $\Omega_0$ approaches unity, $\,p^*_q$ increases for any 
f\/ixed $q$, which implies that a decreasing subset of lens space 
topologies are excluded.

Alternatively, we can ask how f\/ixing the cosmological
parameters results in the exclusion of a given topology as 
a possible shape of our universe. To this end, given a f\/ixed 
value of $r_{-}^{max} $, one can solve the equation 
\be
\label{largeUniv2}
\chi(z_{max}) = r_{-}^{max} \, ,
\ee
which is analogous to (\ref{largeUniv}). 

We recall that if one can be certain that there is no multiple 
images up to a certain depth $z_{max}$ then one can claim that 
$r_-^{max} > \chi_{obs}\,$, and therefore any topology for which 
$r_-^{max} < \chi_{obs}\,$ can be excluded by such observations. 
Thus, for a f\/ixed catalogue depth $z_{max}\,$, the points 
above the graph of the solution of~(\ref{largeUniv2}) in the plane 
$\varepsilon_{\Lambda}$~--~$\varepsilon_0$ correspond to cosmological
models for which any topology with $r_-^{max}  < \chi_{obs}\,$ 
can be excluded.

As an interesting example of application of the indicator $T_{max}$,
we consider the hyperbolic manifolds and recall the important recent 
mathematical result according to which any closed orientable 
hyperbolic 3--manifold contains a ball of radius 
$r_0 = 0.24746$~\cite{Przeworski}.
It therefore follows that the absolute lower bound to $r_{-}^{max}$ 
for any compact hyperbolic manifold is given by $r_0$. 
We solved  Eq.~(\ref{largeUniv2}) for $r_-^{max}=r_0$
and with catalogue depths $z_{max} = 6$ and $3000\,$,
and the results are shown in Fig.~3. This f\/igure shows  
that the allowed observational hyperbolic range of the 
cosmological parameters given by bounds ({\bf ii})
intersects the lowest solution curve, corresponding to
the catalogue with $z_{max} =3000$.
Thus, given the current observational bounds ({\bf ii}), 
if no repeated patterns exist up to $z_{max}=3000$,
there are families of hyperbolic manifolds that  
can be excluded as the 
shape of our universe.%
\footnote{Clearly we are assuming here that the universe is
multiply connected, and no multiple images exist. These
multiply connected universes are therefore indistinguishable 
(using pattern repetition) from simply-connected universes 
with the same covering space, equal radius, and identical 
distribution of cosmic sources.
In these cases the scale of multiply-connectedness is
greater than the observations depth $\chi_{obs}(z_{max})$ 
and no sign of multiply-connectedness will arise.}
As concrete examples note that in such cases the topology 
of the f\/irst ten closed orientable hyperbolic manifolds 
with smallest volumes  given in Table~\ref{Tb:HW-Census} 
would not be excluded, as they all possess values of 
$r_{-}^{max} \ge 0.519162$, and therefore the corresponding 
solution curves would lie substantially above the hyperbolic
range of the observational bounds ({\bf ii}).

Interestingly Fig.~3 also shows that if only quasars are used 
(redshift up to $z_{max} = 6$) and no multiple images exist,
then no hyperbolic manifolds can be excluded as the shape of 
our universe, as the corresponding solution curve lies above the 
current bounds ({\bf ii}) on the cosmological parameters. 

The above considerations provide examples
of the importance of considering families of manifolds
(topologies), rather than arbitrary individual examples,
in looking for the shape of our universe in the inf\/inite 
set of distinct possible topologies.

\section{Final Remarks}  \label{Concl}

We have made a detailed study of the question of detectability of
the cosmic topology in nearly f\/lat universes ($\Omega_0 \sim 1$),
which are favoured by recent cosmological observations. 
Most studies so far have concentrated on individual manifolds. 
Given the inf\/inite number of theoretically possible topologies, 
we have instead concentrated on how to employ current observations 
and a number of indicators in order to f\/ind 
families of possible undetectable manifold (topologies) as well
as families of manifolds (topologies) that can be excluded.

We have found that as $\Omega_0 \to 1$, increasing families
of possible manifolds (topologies) become undetectable observationally.
In this sense the topology of the universe can be said to 
become more dif\/f\/icult to detect through 
observations of multiple images of either 
cosmic objects or spots of microwave background radiation.
We have also found that for any given manifold $M$ with non-zero 
curvature, there are values of $|\Omega_0 -1|$ below
which its topology is undetectable (using pattern repetition)
for any mix of $\Omega_{m0}$ and $\Omega_{\Lambda 0}$.

We have made a detailed study of the constraints
that the most recent estimates of the cosmological
parameters place on the detectable and allowed topologies.
Considering concrete examples of both spherical and
hyperbolic manifolds, we f\/ind that, given the present 
observaional bounds on cosmological parameters,
there are families of both hyperbolic and spherical 
manifolds that remain undetectable and
families that can be excluded as the shape of our universe.
We also demonstrate the importance 
of considering families of 
possible manifolds (topologies), 
rather than arbitrary individual examples,
in search strategies for the detection of
the shape of our universe.

Finally we note that even though the precise values of 
the recent bounds on the density parameter used in this
paper are likely to be modif\/ied by future observations, 
the closeness  $\Omega_0 \sim 1 $ is expected to be 
conf\/irmed.
We have chosen these bounds in this paper as concrete 
examples of how recent observations may be employed in 
order to constrain the topology of the universe, and 
clearly similar procedures can be used for any 
modif\/ied future bounds on cosmological parameters.

\vspace{3mm}
\section*{Acknowledgments}
We are grateful to Jef\/f Weeks for his expert advice on 
topology and on the SnapPea program.
Our knowledge of hyperbolic manifold was greatly enhanced
by his very kind and useful correspondence and comments. 
We also thank Dr A. Przeworski for helpful correspondence.
Finally we thank FAPERJ and CNPq for the grants under which 
this work was carried out.

\vspace{3mm}
\section*{Figure captions}

\begin{description}
\item[Figure 1.] 
The behaviour of the horizon radius $\chi_{hor}$
in units of curvature radius,
for FLRW models with dust and 
cosmological constant, given by Eq.~(\ref{horLambda}),
as a function of the cosmological density
parameters $\Omega_{\Lambda}$ and $\Omega_m\,$.
These f\/igures (1a \& 1b) show clearly the rapid way
$\chi_{hor}$ (as well as $\chi_{obs}$) falls off to zero
for nearly f\/lat (hyperbolic or elliptic) universes,
as $\Omega_0 = \,\Omega_{m0} + \Omega_{\Lambda 0} \,\,\to 1$.
In both f\/igures the vertical axes represent
$\chi_{hor}$, while the horizontal axes are given respectively
(and anti--clockwise), by $\Omega_{\Lambda}$ and $\Omega_m$ (in Fig.~1a)
and $\Omega_m$ and $\Omega_{\Lambda}$ (in Fig.~1b).

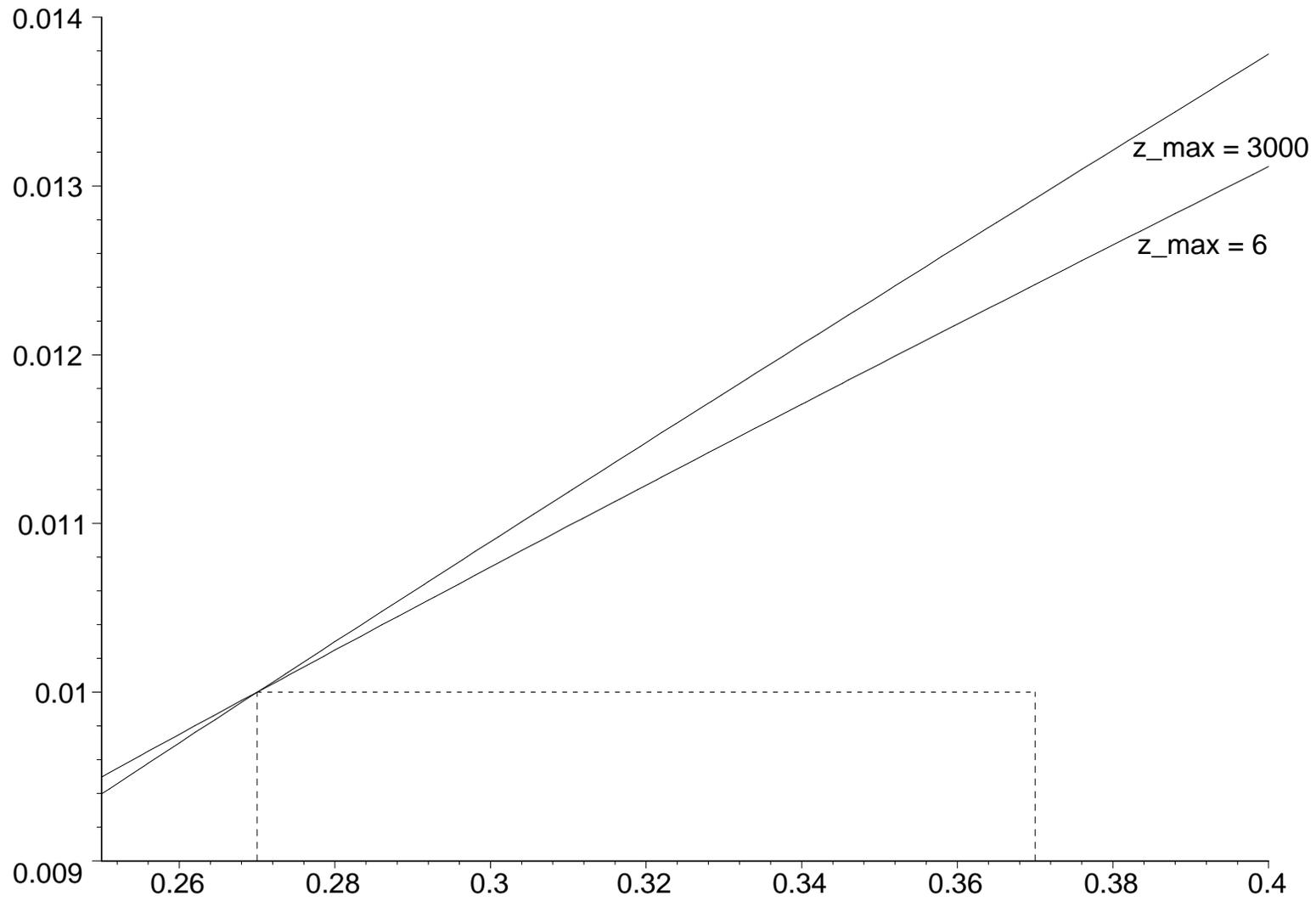
\item[Figure 2.]
The solutions of Eq.~(\ref{epseq}) as plots of
$\epsilon_0 =1-\Omega_{\Lambda 0}$ (vertical  axis) versus 
$\epsilon_\Lambda =1-\Omega_{\Lambda 0}\,$ (horizontal axis),
with $r_{inj}$ taken as the
largest values of the $\chi_{obs}$ in the range of cosmological
parameters given by the bounds ({\bf ii}). The upper and the lower
solutions curves correspond to the catalogues
with $z_{max}=3000\,$ and $z_{max}=6$, respectively. 
Note that the allowed hyperbolic region of the parameter space
given by ($\varepsilon_0 \in (0, 0.01]$ \& 
$\varepsilon_{\Lambda} \in [0.27,0.37]$), and represented by 
the dashed rectangular box, lies below both curves, showing 
that using quasars up to $z_{max}=6\,$ nearly f\/lat FLRW hyperbolic
universes, with the density parameters in this region, will
have undetectable topologies, if their corresponding inradii 
satisfy $r_{inj} \geq  0.20125\,$. 
Similarly, using CMBR, with $z_{max}=3000\,$,
the topology of nearly f\/lat hyperbolic FLRW universes with 
$r_{inj} \geq 0.34211$ will be undetectable.

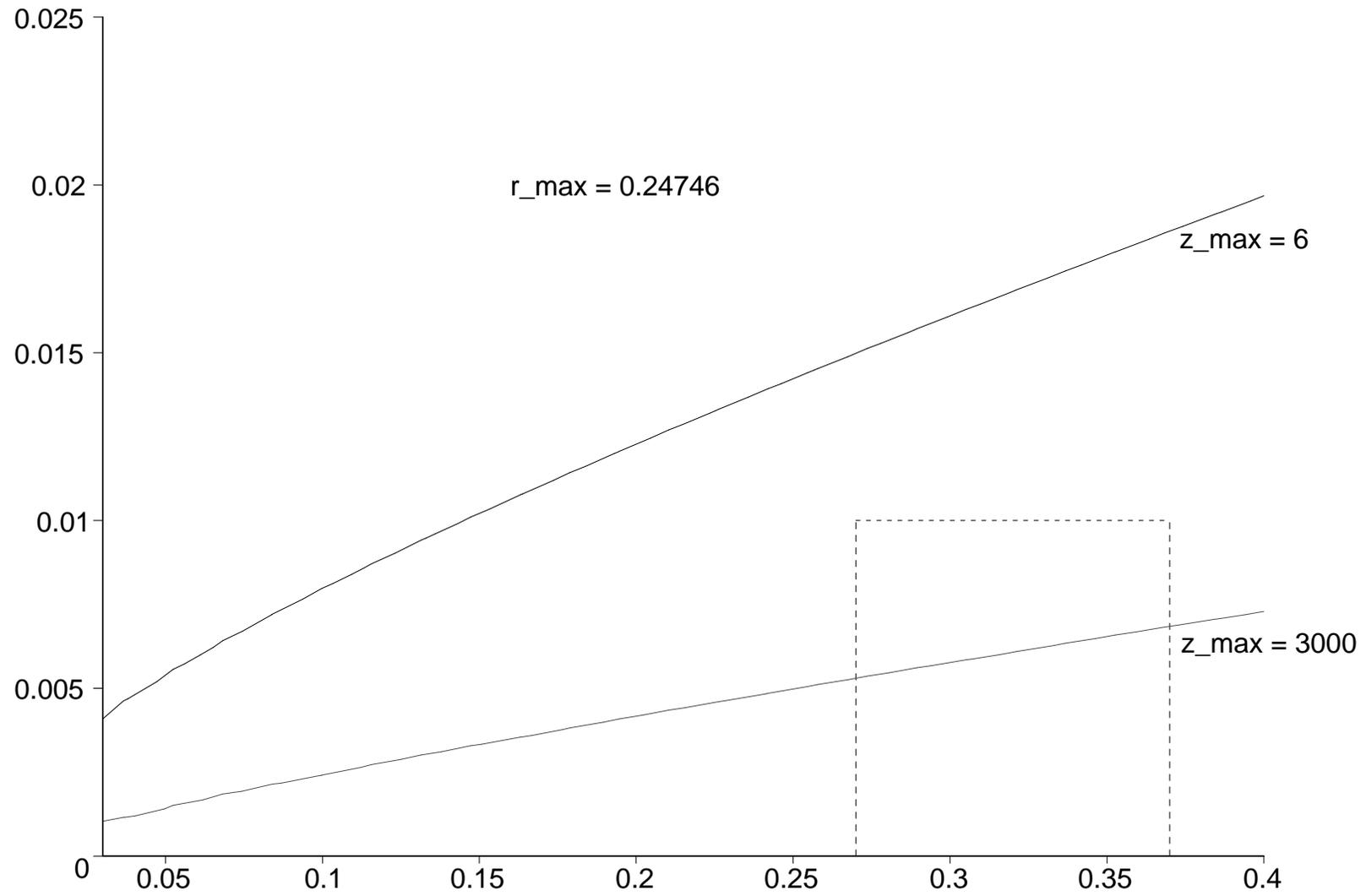
\item[Figure 3.]
The solutions of Eq.~(\ref{largeUniv2}) for $r_-^{max} = 0.24746$, 
as plots of $\epsilon_0$ (vertical axis) versus $\epsilon_\Lambda $ 
(horizontal axis),for the catalogues with $z_{max} = 6$ and $z_{max}=3000$.
As can be seen, the allowed observational hyperbolic range of the 
cosmological parameters given by bounds ({\bf ii}), and represented by 
the dashed rectangular box, intersects the lowest solution curve, 
corresponding to the catalogue with $z_{max} =3000$, thus showing that given 
the current observational bounds, there are families of hyperbolic manifolds 
that remain undetectable and families that can be excluded as the shape of
our universe, if there are no repeated patterns up to $z_{max}=3000$. 
On the other hand if quasars ($z_{max} = 6$) are used, then no hyperbolic 
manifolds can be excluded as the shape of our universe, if there are no 
repeated patterns up to that depth, as the corresponding solution curve 
lies above the dashed rectangle.
\end{description}


%
%
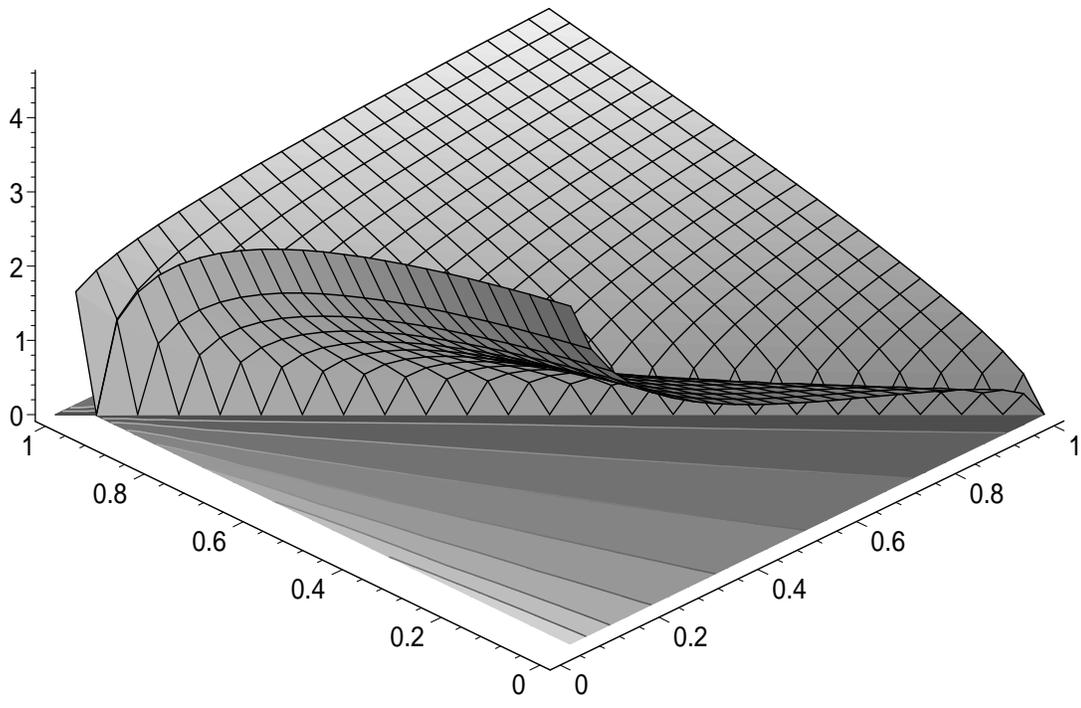
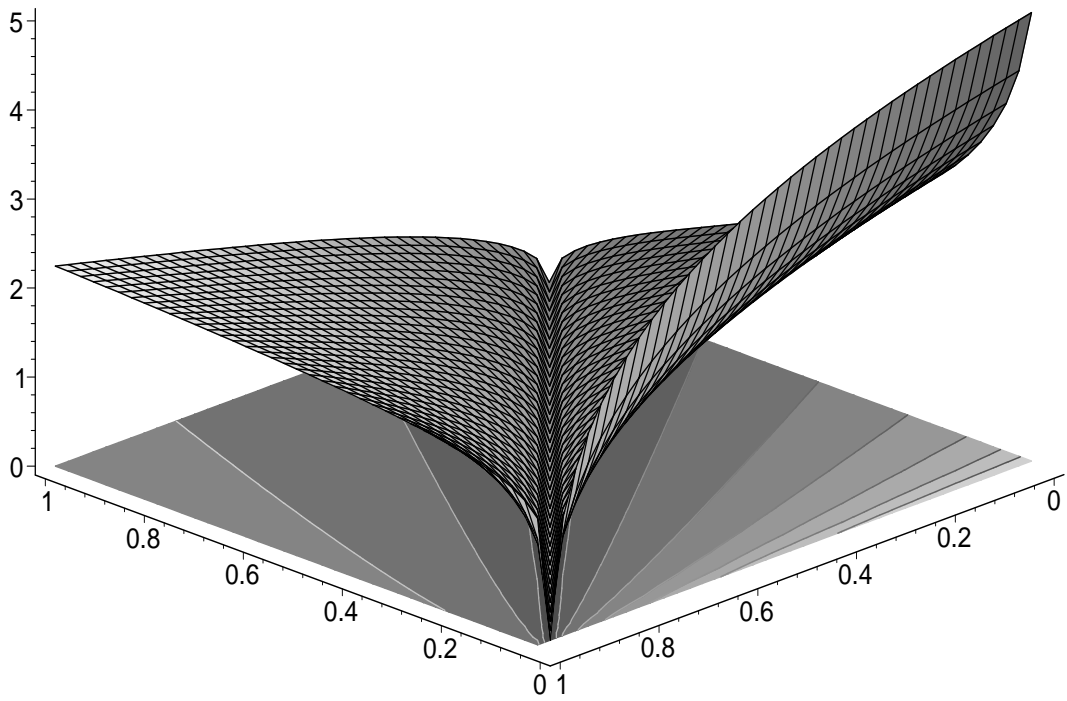
\end{document}